\documentclass{rspublic}
\usepackage{epsfig}
\usepackage{latexsym}
\begin{document}
\title[Thermohydrodynamics of lamellar fluids]{Thermal and hydrodynamic
effects in the ordering of lamellar fluids}
\author[G. Gonnella and others]{G. Gonnella$^1$, A. Lamura$^2$
and A. Tiribocchi$^1$}
 \affiliation{$^1$Dipartimento di
Fisica, Universit\`{a} di Bari,
 {\it and} INFN, Sezione di Bari,
Via Amendola 173, 70126 Bari, Italy \\
$^2$Istituto Applicazioni Calcolo, CNR,
Via Amendola 122/D, 70126 Bari, Italy}
\label{firstpage}
\maketitle
\begin{abstract}{Thermal effects; hydrodynamics;
lattice Boltzmann method; lamellar order} 
Phase separation in a
complex fluid  with lamellar order has been studied in the case of
cold thermal fronts propagating diffusively from external walls. The
velocity hydrodynamic modes are taken into account by coupling the
convection-diffusion equation for the order parameter to a
generalised Navier-Stokes equation. The dynamical  equations are
simulated by implementing a hybrid method based on a lattice
Boltzmann algorithm coupled to finite difference schemes.
Simulations show that the ordering process occurs  with morphologies
depending on the speed of the thermal fronts or, equivalently, on
the value of the thermal conductivity $\xi$. At large value of
$\xi$, as in instantaneous quenching, the system is frozen in
entangled configurations at high viscosity while consists of grains
with well ordered lamellae at low viscosity. By decreasing the value
of $\xi$, a regime with very ordered  lamellae parallel to the
thermal fronts is found. At very low values of $\xi$ the preferred
orientation is perpendicular to the walls  in $d=2$, while
perpendicular order is lost moving  far from the walls in $d=3$.
\end{abstract}

\section{Introduction}
Ordering in complex fluids can be a slow process.  Thermodynamic
equilibrium is often  characterised by the presence of organised
structures at mesoscopic scales (Larson 1999). 
However,  the dynamical formation
of topological defects can make difficult to reach the true
equilibrium state (Boyer \& Vi\~{n}als 2002, Benzi {\it et al.} 2009). 
Since there are physically relevant
properties related to the degree of order in the system, it can be
crucial to control the ordering process by proper driving.

In this paper we will consider fluids with lamellar order, such as
symmetric block copolymer melts (Fredrickson \& Bates 1996). For
recent applications of these systems in micro- and nano-technologies
one can  for example see  the work by Man {\it et al.} (2010) where
the relevance of having  long-range order is discussed. A transient
macroscopic sample of  a copolymer melt usually consists of
different oriented lamellar domains (grains) with  defects like
dislocations and disclinations. The interplay
between the mesoscopic structures and the local velocity field is
essential for the evolution of defects (Gonnella {\it et
al.} 1997). Hydrodynamic modes favour the disentanglement of
intertwined patterns but  simulations have shown how extended
defects between different grains remain at late times slowing down
the dynamics of ordering (Xu {\it et al.} 2004, 2005).

Mechanical driving like shear can favour ordering since interfaces
generally prefer to be aligned with the flow (Larson 1999). In lamellar
systems under shear, however, different stable orientations are
possible (Fredrickson 1994) and in the ordering process  (Cates \&
Milner 1989, Corberi {\it et al.} 2002)  the shear by itself can
cause the creation of defects due to the dilatation or compression of
the layers (Kumaran {\it et al.} 2001, Lu {\it et al.} 2008).  More
complex phenomena like shear banding have been also observed (Xu
{\it et al.} 2006).

A different controlling procedure  could be thermal driving.
Recently, in simple binary mixtures, it has been shown than in
thermally controlled quenches with   cold fronts propagating into
the system at finite speed, interfaces have a preferred orientation
with respect to the fronts (Furukawa 1992; Foard \& Wagner 2009;
Gonnella {\it et al.} 2010). As a result, ordered patterns are
obtained. Applications to polymer systems have been  considered for
example by Voit {\it et al.} (2005). In this work we want to
investigate how a thermally controlled quench could be used to
produce  well ordered lamellar systems.

We consider a system described in the continuum by the Navier-Stokes
and convection-diffusion equations with the thermodynamics of the
lamellar phase encoded in a free-energy functional. The temperature
evolves  by following a standard diffusion equation with the system
between two walls at a temperature below  the critical value. In
order to solve the dynamical equations we apply   a hybrid method
with a lattice Boltzmann algorithm for solving  the Navier-Stokes
equations and a finite difference scheme for the
convection-diffusion equation. The model and the method will be
described in the next section while the  numerical results will be
presented  Section III. Few conclusions will complete the paper.

\section{The model}
We  consider a binary mixture having total density $n$ and density
difference  $\varphi$,  the order parameter, between the two
components.  The equilibrium properties are described by the free
energy (Brazovskii 1975)
\begin{equation}
{\cal F}=\int
\left[nT+nT\ln(nT)+\frac{a}{2} \varphi^2+ \frac{b}{4} \varphi^4
+ \frac{\kappa}{2} (\nabla \varphi)^2 + \frac{c}{2} (\nabla^2 \varphi)^2
\right]\rd{{\bf r}} .
\label{fren}
\end{equation}
The first two terms depending on  $n$ and on the temperature $T$ (the
Boltzmann constant is assumed to be one) give the ideal
gas pressure which does not affect the phase behaviour. We
consider $b, c > 0$ to ensure stability. The parameter $a$ can be
thought of as a reduced temperature
\begin{equation}
a=\frac{T-T_c}{T_c} 
\label{aterm}
\end{equation}
and is the only one depending on temperature
in the terms depending on $\varphi$.
For temperatures above the critical value $T_c$ the fluid is
disordered. For $T<T_c$ and $\kappa >0$ there is the coexistence of two
homogeneous phases with equilibrium values $\varphi=\sqrt{-a/b}$.
Negative values of $\kappa$ favour the formation of interfaces,
while a reduction of $\kappa$ can induce a transition into a
lamellar phase. By using a single mode approximation with profiles
like $A \sin k_0 x$ in the direction normal to lamellae, one finds a
transition, when $|a|=b$, at $a \simeq -1.11 \kappa^2/c$ with
$k_0=\sqrt{-\kappa/2 c}$ and $A^2=4(1+\kappa^2/4cb)/3$ (Xu {\it et
al.} 2005).
The parameters of this model 
have been mapped on those of copolymer systems by Binder (1995).

The evolution of the system is described by a set of coupled partial
differential equations:
\begin{equation}
\partial_t n + \partial_{\alpha}(n u_{\alpha})=0 ,
\label{mass}
\end{equation}
\begin{equation}
n (\partial_t u_{\alpha} + u_{\beta} \partial_{\beta} u_{\alpha})=
-\partial_{\beta} P_{\alpha \beta}
+ \partial_{\beta} \left [ \eta (\partial_{\alpha} u_{\beta}+
\partial_{\beta} u_{\alpha})+(\zeta-2 \eta/d)\delta_{\alpha \beta}
\partial_{\gamma} u_{\gamma} \right ] ,
\label{momentum}
\end{equation}
\begin{equation}
\partial_t \varphi+\partial_{\alpha} (\varphi
u_{\alpha})=\Gamma \nabla^2 \mu ,
\label{massdiff}
\end{equation}
where ${\bf u}$ is the local velocity of the
fluid, $\eta$ the shear viscosity, $\zeta$ the bulk viscosity,
$d$ the space dimension, and $\Gamma$ the mobility coefficient. 
We model the quench process by taking the reduced 
temperature $a$ (\ref{aterm}) to be the solution
of a diffusion equation $\partial_t a=\xi \nabla^2 a$
where $\xi$ is the thermal conductivity.
The pressure tensor $P_{\alpha
\beta}$ and the chemical potential $\mu$ can be computed from the
free energy functional (\ref{fren}) and have the following forms
(Gonnella {\it et al.} 1997):
\begin{eqnarray}
&&P_{\alpha \beta}= \Big \{ nT + \frac{a}{2} \varphi^2 +\frac{3b}{4}\varphi^4
-\kappa \Big [ \varphi(\nabla^2 \varphi) + \frac{1}{2} (\nabla \varphi)^2
\Big ] + c \Big [ \varphi (\nabla^2)^2 \varphi+\frac{1}{2}
(\nabla^2 \varphi)^2 \nonumber \\
&&+\partial_{\gamma}\varphi \partial_{\gamma}
(\nabla^2\varphi) \Big ] \Big \} \delta_{\alpha \beta}+
\kappa \partial_{\alpha} \varphi \partial_{\beta} \varphi
- c \Big [  \partial_{\alpha} \varphi \partial_{\beta} (\nabla^2 \varphi)
+ \partial_{\beta} \varphi \partial_{\alpha} (\nabla^2 \varphi) \Big ]
\end{eqnarray}
and
\begin{equation}
\mu=a \varphi + b \varphi^3 -\kappa \nabla^2\varphi+c(\nabla^2)^2\varphi .
\label{chem}
\end{equation}

Equations (\ref{mass})-(\ref{massdiff}) are solved numerically, by using
a mixed approach. A lattice Boltzmann scheme is used for the
continuity and Navier-Stokes equations (\ref{mass}) and
(\ref{momentum}), and a finite-difference approach for equation
(\ref{massdiff}) and for the temperature diffusion equation.
The full study is limited to bi-dimensional systems while the
equations for $\varphi$ not coupled to hydrodynamics
will be solved also  in $d=3$.
In the case of a simple fluid (Benzi {\it et al.} 1992, Chen \&
Doolen 1998, Succi 2001) the lattice Boltzmann method (LBM) 
is defined in terms of a set of
distribution functions, $f_i({\bf r},t)$, located in each lattice
site ${\bf r}$ at each time $t$, and of a set of nine velocity
vectors ${\bf e}_i$, defined on a square lattice, having moduli
$|{\bf e}_i|=0, c, \sqrt{2} c$ with $c=\Delta x/\Delta t$ being
$\Delta x$ and $\Delta t$ the lattice and the time step,
respectively. The distribution functions evolve according to a
single relaxation time Boltzmann equation (Bhatnagar {\it et al.}
1954)
\begin{equation}
f_i({\bf r}+{\bf e}_i\Delta t,t+\Delta t)-f_i({\bf r},t)=-\frac{\Delta t}
{\tau}[f_i({\bf r},t)-f_i^{eq}({\bf r},t)]+\Delta tF_i({\bf r},t),
\label{evoleqn}
\end{equation}
where $\tau$ is the relaxation time, $f_i^{eq}$ are the equilibrium
distribution functions and $F_i$ are the forcing terms to be properly
fixed.

The physical quantities, that is the total density $n$ and the fluid
momentum $n {\bf u}$, are determined by the relations
\begin{equation}\label{moment}
n=\sum_if_i , \hspace{1.3cm} n{\bf u}=\sum_if_i{\bf e}_i
+ \frac{1}{2}{\bf F}\Delta t,
\end{equation}
with ${\bf F}$ the force density acting on the fluid. The
equilibrium distribution functions $f_i^{eq}$ are given by the
standard second order expansion in the fluid velocity ${\bf u}$ of
the Maxwell-Boltzmann distribution function (Qian {\it et al.}
1992).

The forcing term in equation
(\ref{evoleqn}) is expressed as a second order expansion
in the lattice vector velocities (Guo {\it et al.} 2002)
\begin{equation}
F_i=\left(1-\frac{\Delta t}{2\tau}\right)\omega_i\left[
\frac{{\bf e}_i-{\bf u}}{c^2_s}+\frac{{\bf e}_i\cdot{\bf u}}{c^4_s}
{\bf e}_i\right]\cdot {\bf F} ,
\label{latticeforceterm}
\end{equation}
where $c_s=c/\sqrt{3}$ is the speed of sound and $\omega_i=4/9,
1/9, 1/36$ in correspondence of the lattice directions with $|{\bf
e}_i|=0, c, \sqrt{2} c$, respectively. The force ${\bf F}$ has to
have the following expression
\begin{equation}
F_{\beta}=\partial_{\alpha}(nc_s^2\delta_{\alpha \beta}-P_{\alpha\beta})
\label{ff}
\end{equation}
in order to recover equation (\ref{momentum}) where it results
\begin{equation}
\zeta=\eta = n c_s^2 \Delta t\left(\frac{\tau}{\Delta t} -
\frac{1}{2}\right).
\end{equation}
The first term on the r.h.s. of equation (\ref{ff}) allows to cancel out
the athermal ideal gas pressure $nc_s^2$ of the LBM. 
We verified that the Mach number 
$Ma=|{\bf u}|_{max}/c_s$, where
$|{\bf u}|_{max}$ is the maximum value of the fluid velocity
during evolution, stays always smaller than $0.1$ 
so that 
the fluid results practically incompressible with $n \simeq 1$. 

The convection-diffusion (\ref{massdiff}) and the temperature
equations are solved by using a finite-difference scheme,
 which is described with details  in the paper by Tiribocchi {\it
et al.} (2009). The function $\varphi({\bf r},t)$ is defined on the
same sites of the LBM with the same space and time steps. We update
$\varphi$ using an explicit Euler scheme (Strikwerda 1989). The
spatial differential operators are calculated by a second-order
finite-difference scheme.

Periodic boundary conditions are used along the $x$-direction (and
$y$-direction in $d=3$) and flat walls are placed at the lower and
upper rows of the lattice along the $y$-direction ($z$-direction
in $d=3$), where the
temperature is kept at fixed value $T_w$.
Moreover, neutral wetting condition
for the order parameter $\varphi$ is implemented at walls imposing
that ${\bf a}\cdot{\bf \nabla}\varphi|_{wall}=0$, being ${\bf a}$ an
inward normal unit vector to the boundaries. To guarantee exact
conservation of the order parameter $\varphi$ we also require that
${\bf a}\cdot{\bf \nabla}(\nabla^2\varphi)|_{wall}= {\bf a}\cdot{\bf
\nabla}[(\nabla^2)^2\varphi]|_{wall}=0$. The bounce-back rule
(Lavallee {\it et al.} 1991, Desplat {\it et al.} 2001) for the LBM
is adopted at walls supplemented by the prescription for density
exact conservation introduced by Lamura \& Gonnella (2001).

\section{Results and discussion}
In this section we will present the results of our simulations of
equations (\ref{mass})-(\ref{massdiff}). 
We used  the  set of parameters $b=0.1,
\kappa=-c=-0.03, T_c = 0.005$, and $\Gamma=0.1$ corresponding to the
lamellar phase in the phase diagram. Lattices had sizes $256 \times
256$ in $d=2$ and $32 \times 32 \times 256$ in $d=3$. The system,
initially, is in disordered symmetric states with $\varphi$
fluctuating around zero in the interval $[-0.01,0.01]$. The initial
temperature is above $T_c$ with the walls at $T_w=0.9 T_c$.

\begin{figure}[h]
\begin{center}
\epsfig{file=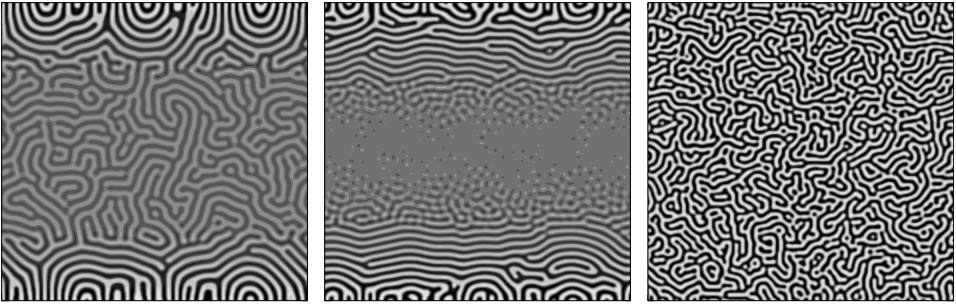,bbllx=62 pt,bblly= 389 pt,bburx= 536 pt,bbury=
542 pt,width=12.6cm}
\end{center}
\caption{Configurations of the concentration field $\varphi$
 without hydrodynamics with thermal conductivity $\xi=10^{-2},
10^{-1}, +\infty$ (from left to right) at times $t=55.7 \times 10^3;
11.7 \times 10^3; 1.95 \times 10^3$, respectively.} 
\label{fig1}
\end{figure}

\begin{figure}[h]
\begin{center}
\epsfig{file=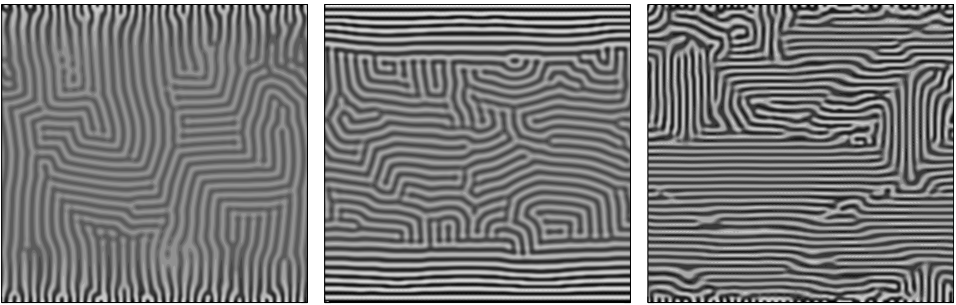,bbllx=62 pt,bblly= 389 pt,bburx= 536 pt,bbury=
542 pt,width=12.6cm}
\end{center}
\caption{Configurations of the concentration field $\varphi$ at low
viscosity ($\eta=1.5$) with thermal conductivity $\xi=10^{-4},
10^{-3}, +\infty$ (from left to right) at times $t=144 \times 10^4;
35.6 \times 10^4; 2.1 \times 10^4$, respectively.} 
\label{fig2}
\end{figure}

\begin{figure}[h]
\begin{center}
\epsfig{file=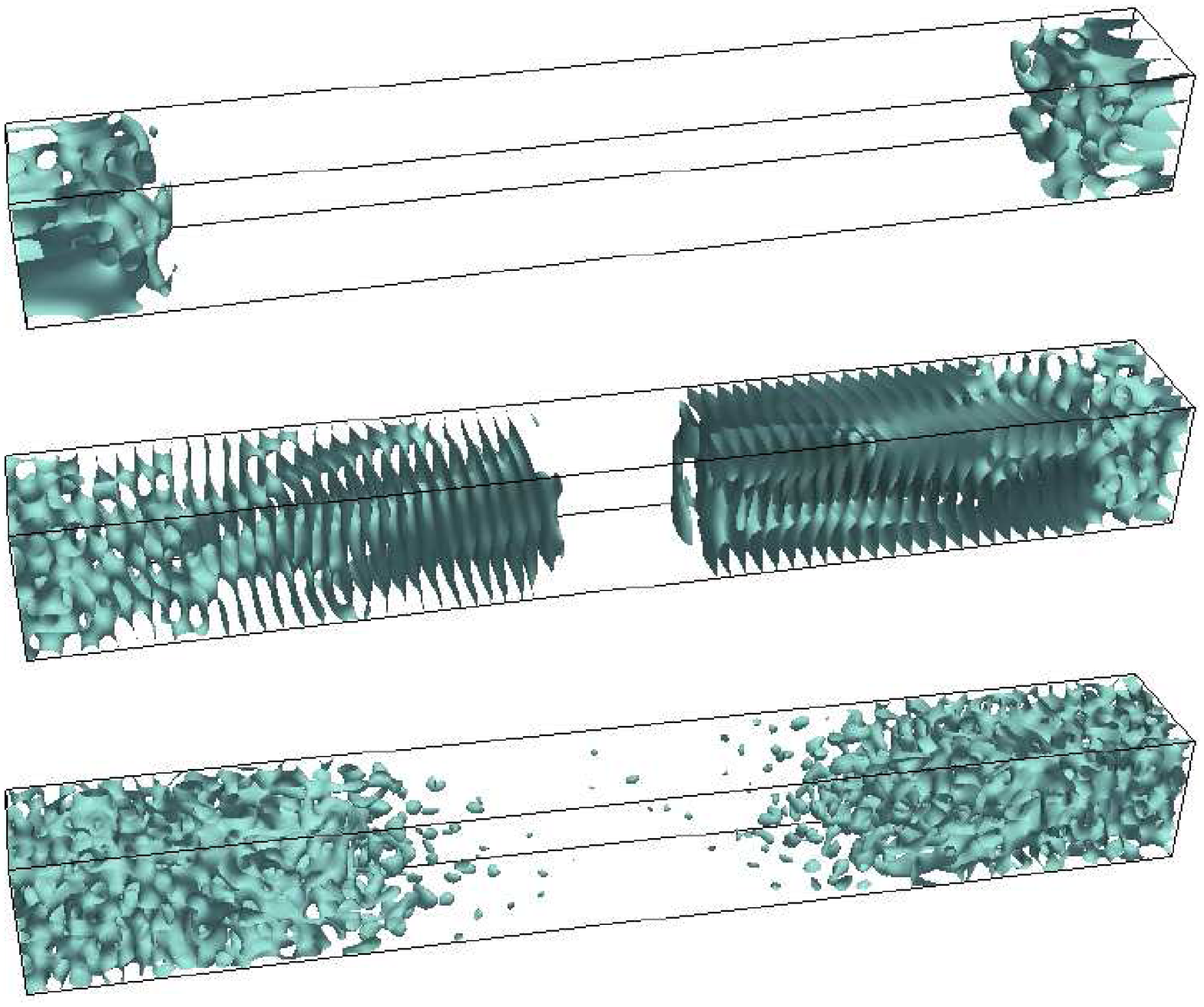,,width=14.6cm}
\end{center}
\caption{Interface configurations in lamellar ordering
without hydrodynamics
with thermal conductivity
$\xi=10^{-2}, 1, 10^2$ (from top to bottom)
at times $t=78 \times 10^2; 15 \times 10^2; 1.1 \times 10^2$, respectively.}
\label{fig3}
\end{figure}

We first show in  figure 1 typical configurations of  systems with very
high viscosity. In this case we   neglected the coupling with the
velocity field by solving equation (\ref{massdiff}) without the convective
term.
Lamellar domains start to form close to the
walls, where the temperature has become   lower than the critical
value, with interfaces perpendicular to the walls in agreement with
the imposed neutral wetting conditions.
 Then, following the temperature fronts, the regions with   
phase separated domains
 will extend  towards the middle of the system. The pattern
morphology, however, depends on the value of the thermal
conductivity or, equivalently, on the speed of the temperature
fronts. At very large $\xi$, the system
remains frozen in intertwined noodle-like configurations  without
long range order, like that on the right of figure 1,  as observed in
other simulations (see, e.g., Gonnella {\it et al.} 1997). However, by
decreasing the value of $\xi$,  in the interval $[0.1,1]$, 
in spite of
the adopted boundary conditions, the  interfaces tend to follow the
temperature fronts, parallel to the walls, and exhibit only  few
defects. The two  fronts separating the regions with ordered
lamellae from the central disordered region will approach each other
joining at the later times of the simulation.
By further decreasing $\xi$, interfaces show the tendency to keep
the perpendicular orientation with respect to the walls.  
This phenomenology has to be compared with that
observed  in simple binary mixtures where at large $\xi$ usual
isotropic phase separation occurs (of course, in binary mixtures,
there is no frustration and the  average size of domains grows by
power law, see, e.g., Bray 1994), while by lowering the values of the thermal
conductivity one observes domains first parallel and then
perpendicular to the cold walls (Tiribocchi {\it et al.}
2010).

Going to a lower viscosity ($\tau=5, \eta=1.5$, see figure 2), at very
large $\xi$, one finds that hydrodynamics favours the
disentanglement of the network pattern (Gonnella {\it et al.} 1997)
but one still observes extended defects between grains of different
oriented lamellae (see e.g. the grain  with  vertical  orientation
on the top-left  region of the system opposed to the large
horizontal underlying domain). One could see in systems larger than
those here considered that these extended defects would slow down
the ordering process at very late times (Xu {\it et al.} 2005).  
At lower thermal
conductivities, in the range $[5 \times 10^{-4}, 5 \times 10^{-3}]$, 
one again finds that
lamellae are well aligned with the thermal fronts. The velocity
helps this ordering and, indeed, very ordered defect-free lamellae
can be observed (compare the central snapshots of figure 1 and figure 2)
in the regions close to the walls. In the central part, 
even if structures can be observed, the local values of the field 
$\varphi$ are nor in equilibrium, and further evolution will show
a completely ordered lamellar state.
Finally, by further decreasing the value of $\xi$,   one sees in the
left snapshot of  figure 2 that hydrodynamics clearly helps  the
tendency of the system to prefer the perpendicular orientation.

We also studied three-dimensional lamellar systems without coupling
to the velocity field. At large values of $\xi$ the system evolves
towards metastable configurations ordered only on short scales,
similarly to what occurs in two-dimensional systems, as it can be
seen in the bottom snapshot of  figure 3. At lower values of $\xi$ 
($0.1 \leq \xi \leq 10$) one can see stacks of lamellae parallel to the
thermal fronts with very few defects (central snapshot of  figure 3),
while, again similarly to the two-dimensional case of figure 1, 
perpendicular
order is lost far from the walls at very low values of $\xi$.

\section{Conclusions}
In this work we have studied the ordering of a lamellar phase where
the temperature cold fronts 
diffusively move from external walls. Phase separation therefore
starts close to the walls but develops in the middle of the systems
with a phenomenology depending on the speed of the thermal fronts.
In an intermediate range of values of thermal conductivity $\xi$,
lamellae appear ordered and parallel to the thermal fronts
while a perpendicular orientation is preferred at very low values of
$\xi$. Hydrodynamics favours these preferred ordering and one can
conclude that quenching with moving temperature fronts are effective
in producing lamellar states with very few defects. A full analysis
of the three-dimensional case will complete this study.

\begin{acknowledgements}
We acknowledge support from CASPUR for ``HPC Grant 2010''.
\end{acknowledgements}

\label{lastpage}
\end{document}